\documentclass[pre,showpacs,preprintnumbers,amsmath,amssymb]{revtex4}

\usepackage{graphicx}
\usepackage{dcolumn}
\usepackage{bm}

\begin{document}

\hoffset = +0.0truein
\voffset = +0.3truein

\preprint{PREPRINT}

\title{Lattice Boltzmann Model for Axisymmetric Multiphase Flows}

\author{Kannan N. Premnath}
 \email{nandha@ecn.purdue.edu}
\author{John Abraham}%
 \email{jabraham@ecn.purdue.edu}
\affiliation{
M.J. Zucrow Labs., School of Mechanical Engineering\\
Purdue University, West Lafayette, IN 47907.
}%

\date{\today}

\begin{abstract}
In this paper, a lattice Boltzmann (LB) model is presented for axisymmetric multiphase flows.
Source terms are added to a two-dimensional standard lattice Boltzmann equation (LBE) for
multiphase flows such that the emergent dynamics can be transformed into the axisymmetric
cylindrical coordinate system. The source terms are temporally and spatially dependent
and represent the axisymmetric contribution of the order parameter of fluid phases and inertial, viscous
and surface tension forces. A model which is effectively explicit and second order is obtained.
This is achieved by taking into account the discrete lattice effects in the Chapman-Enskog
multiscale analysis, so that the macroscopic axisymmetric mass and momentum equations for
multiphase flows are recovered self-consistently. The model is extended to incorporate reduced compressibility
effects. Axisymmetric equilibrium drop formation
and oscillations, breakup and formation of satellite droplets
from viscous liquid cylindrical jets through Rayleigh capillary instability and drop collisions are presented. Comparisons of
the computed results with available data show satisfactory agreement.
\end{abstract}

\maketitle

\section{\label{sec:intro}Introduction}
Fluid flow with interfaces and free surfaces is common in nature and in many engineering
applications. Such interfacial flows which typically involve multiple scales remain a formidable non-linear
problem rich in physics and continue to pose challenges to experimentalists and theoreticians
alike~\cite{eggers97}.  Numerical simulation of multiphase flows is challenging
as the shape and location of the interfaces must be computed in conjunction with the solution
of the flow field~\cite{hyman84,scardovelli99}. Computational methods based on the
lattice Boltzmann equation (LBE) for simulating complex emergent physical phenomena have
attracted much attention in recent years~\cite{chen98,succi02}. The LBE simulates multiphase 
flows by incorporating interfacial physics at scales smaller than macroscopic scales. Phase
segregation and interfacial fluid dynamics can be simulated by incorporating inter-particle
potentials~\cite{shan93,shan94}, concepts based on free energy~\cite{swift95,swift96}
or kinetic theory of dense fluids~\cite{he98,he99,he02}.

The formulation of the standard LBE is based on the Cartesian coordinate system and does not
take into account axial symmetry that may exist. Numerous multiphase
flow situations exist where the fluid dynamics can be approximated as axisymmetric~\cite{sussman96,eggers97}.
Examples include head-on collision of drops, normal drop impingement on solid surfaces and
Rayleigh instability of cylindrical liquid columns. Currently, full three-dimensional (3D)
calculations have to be carried out for problems which may be approximated as axisymmetric~\cite{he99a,inamuro03,premnath04}.
In 3D computations, computational considerations restrict
the numerical resolution that may be employed and the physics may not be well resolved. 
For example, in breakup of drops into satellite droplets the size of the droplets may be such that the 3D
grids may not resolve them. To improve the
computational efficiency of the LBE for axisymmetric multiphase flows, we propose an axisymmetric LB
model in this paper. The approach consists of adding source terms to the two-dimensional (2D) Cartesian LBE
model based on the kinetic theory of dense fluids for multiphase flows~\cite{he98,he99}. This
approach is similar in spirit to the idea proposed in~\cite{halliday01} to solve single-phase axisymmetric
flows. However, multiphase flow problems involve additional complexity as a result of interfacial
physics involved, i.e. the surface tension forces and the need to track the interfaces. In this case,
the accuracy of the numerical discretization of the source terms representing interfacial physics
also becomes an important consideration.

This paper is organized as follows. In Section \ref{sec:axismodel}, the axisymmetric LBE multiphase model is
described. Then, in Section \ref{sec:axismodelc}, its extension to simulate axisymmetric multiphase flows
with reduced compressibility effects is described. The computational methodology adopted is also discussed in this
section. In Section \ref{sec:results}, the axisymmetric model is applied to benchmark problems to evaluate its accuracy.
Finally, the paper closes with summary in Section \ref{sec:summary}.

\section{\label{sec:axismodel}Axisymmetric LBE Multiphase Flow Model}
To simulate axisymmetric multiphase flows, axisymmetric contributions of the order parameter, and
inertial, viscous and surface tension forces may be introduced to the standard 2D LBE. The source
terms, which will be shown to be spatially and temporally dependent, are determined by performing
a Chapman-Enskog multiscale analysis in such a way that the macroscopic mass and momentum
equations for multiphase flows are recovered self-consistently. The introduction of source terms
makes it necessary to calculate additional spatial gradients when compared to those in the standard
LBE. While this approach is developed for a specific LBE multiphase flow model based on kinetic theory
of dense fluids~\cite{he98,he99}, it can be readily extended to other LBE multiphase flow models.

The governing continuum equations of isothermal multiphase flow~\cite{nadiga96,zou99} in the cylindrical
coordinate system when the axisymmetric assumption is employed are
\begin{equation}
\partial_t \rho + \frac{1}{r} \partial_r \left( \rho r u_r  \right) +
\partial_z \left( \rho u_z \right) = 0,
\label{eq:axiscont}
\end{equation}
\begin{equation}
\rho\left(  \partial_t u_r + u_r \partial_r u_r + u_z \partial_z u_r  \right)=
-\partial_r P + F_{s,r}+F_{ext,r}+\frac{1}{r}\partial_r \left( r \Pi_{rr} \right)+
\partial_z \left( \Pi_{rz} \right),
\label{eq:axismomr}
\end{equation}
\begin{equation}
\rho\left(  \partial_t u_z + u_r \partial_r u_z + u_z \partial_z u_z  \right)=
-\partial_z P + F_{s,z}+F_{ext,z}+\frac{1}{r}\partial_r \left( r \Pi_{zr} \right)+
\partial_z \left( \Pi_{zz} \right),
\label{eq:axismomz}
\end{equation}
where $\rho$ is the density and $u_r$ and $u_z$ are the radial and axial components of velocity. 
These equations are derived from kinetic theory that incorporates intermolecular interactions forces
which are modeled as a function of density following the work of van der Waals~\cite{rowlinson}. The exclusion volume
effect of Enskog~\cite{chapman} is also incorporated to account for increase in collision probability due to the
increase in the density of non-ideal fluids. These features naturally give rise to surface tension and phase
segregation effects. The other
variables which appear in the above equations will now be described. $\Pi_{rr}$, $\Pi_{rz}$, $\Pi_{zz}$ are
the components of the viscous stress tensor and are given by
\begin{eqnarray}
\Pi_{rr}&=&2\mu \partial_r u_r, \\
\Pi_{rz}&=& \Pi_{zr}=\mu \left( \partial_z u_r + \partial_r u_z  \right),\\
\Pi_{zz}&=&2\mu \partial_z u_z,
\end{eqnarray}
where $\mu$ is the dynamic viscosity. $F_{s,r}$ and $F_{s,z}$ are the axial and radial components respectively
of the surface tension force, which are given by~\cite{zou99}
\begin{eqnarray}
F_{s,r}&=& \kappa \rho \partial_r \left[\frac{1}{r} \partial_r (r\partial_r\rho)+\partial_z(\partial_z\rho) \right],
\label{eq:surfr}\\
F_{s,z}&=& \kappa \rho \partial_z \left[\frac{1}{r} \partial_r (r\partial_r\rho)+\partial_z(\partial_z\rho) \right],
\label{eq:surfz}
\end{eqnarray}
where $\kappa$ controls the strength of the surface tension force. This parameter is
related to the surface tension of the fluid, $\sigma$, through the density gradient across the interface by the equation~\cite{evans79}
\begin{equation}
\sigma = \kappa \int \left( \frac{\partial \rho}{\partial n} \right)^2 dn.
\label{eq:sigmakappa}
\end{equation}
Thus, the surface tension is a function of both the parameter $\kappa$ and the density profile across the interface.
The terms $F_{ext,r}$ and $F_{ext,z}$ in Eqs. (\ref{eq:axismomr}) and (\ref{eq:axismomz}) respectively
are the radial and axial components of external forces such as gravity.

The pressure, $P$, is related to density through the Carnahan-Starling-van der Waals equation of state (EOS)~\cite{carnahan69}
\begin{equation}
P=\rho R T \left\{ \frac{1+\gamma+\gamma^2-\gamma^3}{(1-\gamma)^3}  \right\} - a\rho^2,
\label{eq:axiseos}
\end{equation}
where $\gamma=b\rho/4$. The parameter $a$ is related to the intermolecular pair-wise potential and $b$
to the effective diameter of the molecule, $d$, and the mass of a single molecule, $m$, by
$b=2\pi d^3/3m$. $R$ is a gas constant and $T$ is the temperature. 
The Carnahan-Starling EOS has a \emph{supernodal} $P-1/\rho-T$ curve, i.e., $dP/d\rho<0$, for certain range of values of $\rho$,
when the state fluid temperature is below its critical value. This unstable part of the curve is the driving
mechanism responsible for keeping the phases of fluids segregated and for maintaining a self-generated
sharp interface.

We now modify the standard LBE in such a way that it effectively yields the axisymmetric multiphase flow equations, 
Eqs. (\ref{eq:axiscont})-
(\ref{eq:axiseos}), in a self-consistent way. To facilitate this,
we employ  the following coordinate transformation, illustrated in Fig.~\ref{fig:schemaxis}, which allows the governing equations to be
represented in a Cartesian-like coordinate system, i.e. $(x,y)$:
\begin{figure*}
\includegraphics{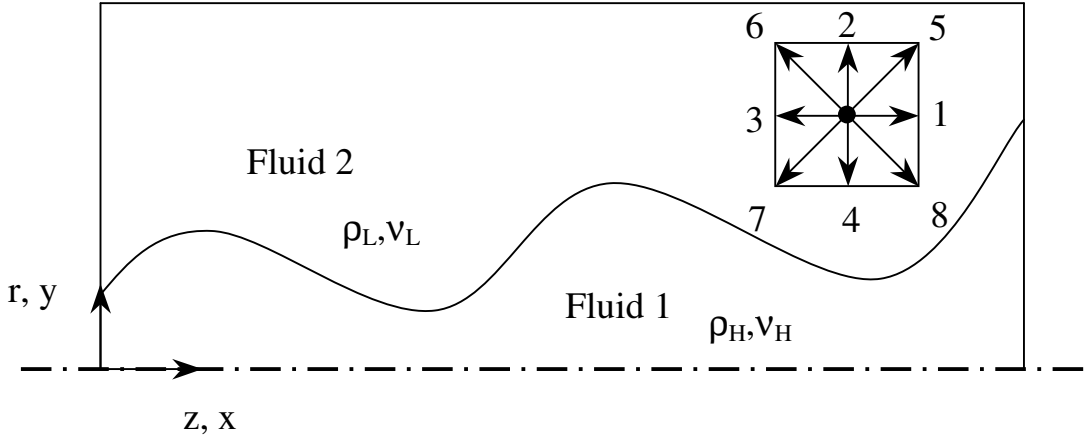}
\caption{\label{fig:schemaxis}
Schematic of arrangement of coordinate system in axisymmetric multiphase flow ($(r,z)$ and $(y,x)$
	 coordinate directions are shown).}
\end{figure*}
\begin{equation}
(r,z) \rightarrow (y,x),
\end{equation}
\begin{equation}
(u_r,u_z) \rightarrow (u_y,u_x).
\end{equation}
Assuming summation convention for repeated subscript indices, Eqs. (\ref{eq:axiscont})-(\ref{eq:surfz}) may be
transformed to
\begin{equation}
\partial_t \rho + \partial_k \left( \rho u_k \right)=-\frac{\rho u_y}{y},
\label{eq:axiscont1}
\end{equation}
\begin{equation}
\rho \left( \partial_t u_i+ u_k\partial_k u_i \right)=-\partial_i P + F_{s,i}+F_{ext,i}+
\partial_k \left[ \mu\left( \partial_k u_i+\partial_i u_k \right) \right]+F_{ax,i},
\label{eq:axismom1}
\end{equation}
where
\begin{equation}
F_{s,i}=\kappa \rho \partial_i \nabla^2 \rho
\label{eq:forcemp}
\end{equation}
and $i,j,k\in \left\{ x,y  \right\}$. The right hand side (RHS) in Eq. (\ref{eq:axiscont1}), $-\rho u_y/y$, is the
additional term in the continuity equation that arises from axisymmetry. The corresponding term
for the momentum equation, Eq. (\ref{eq:axismom1}), is
\begin{equation}
F_{ax,i}=\frac{\mu}{y}\left[\partial_y u_i+\partial_i u_y  \right]+
\kappa \rho \partial_i \left( \frac{1}{y}\partial_y \rho \right).
\end{equation}

To recover Eqs. (\ref{eq:axiscont1}) and (\ref{eq:axismom1}), we introduce two additional source terms,
$S_{\alpha}^{'}$ and $S_{\alpha}^{''}$, to the standard 2D Cartesian LBE which has $\Omega_{\alpha}$ as its collision term and
a source term for the internal and external forces, $S_{\alpha}$. These unknown additional terms, representing the
axisymmetric mass and momentum contributions respectively, are to be determined so that the macroscopic
behavior of the proposed LBE corresponds to axisymmetric multiphase flow. Thus, we propose the
following LBE
\begin{eqnarray}
f_{\alpha}( \mbox{\boldmath$x$}+\mbox{\boldmath$e$}_{\alpha}\delta_t,t+\delta_t )-f_{\alpha}( \mbox{\boldmath$x$},t )&=&
\frac{1}{2}\left[\Omega_{\alpha}|_{(x,t)}+
		 \Omega_{\alpha}|_{(x+e_{\alpha}\delta_t,t+\delta_t)}
\right]+\nonumber\\
& &
\frac{1}{2}\left[S_{\alpha}|_{(x,t)}+
S_{\alpha}|_{(x+e_{\alpha}\delta_t,t+\delta_t)}
\right]\delta_t+\nonumber\\
& &
\frac{1}{2}\left[S_{\alpha}^{'}|_{(x,t)}+
S_{\alpha}^{'}|_{(x+e_{\alpha}\delta_t,t+\delta_t)}
\right]\delta_t+\nonumber\\
& &
\frac{1}{2}\left[S_{\alpha}^{''}|_{(x,t)}+
S_{\alpha}^{''}|_{(x+e_{\alpha}\delta_t,t+\delta_t)}
\right]\delta_t,
\label{eq:axislbe}
\end{eqnarray}
where $f_{\alpha}$ is the discrete single-particle distribution function,  corresponding to the particle velocity,
$\mbox{\boldmath$e$}_{\alpha}$, where $\alpha$ is the velocity direction. The Cartesian component of the particle velocity, $c$,
is given by $c=\delta_x/\delta_t$, where $\delta_x$ is the lattice spacing and $\delta_t$ is the time step corresponding to the
two-dimensional, nine-velocity model(D2Q9)~\cite{qian92} shown in Fig.~\ref{fig:schemaxis}.
Here, the collision term is given by the BGK approximation~\cite{bhatnagar54}
\begin{equation}
\Omega_{\alpha}=-\frac{f_{\alpha}-f_{\alpha}^{eq}}{\tau}, \quad \tau=\frac{\lambda}{\delta_t},
\end{equation}
where $\lambda$ is the relaxation time due to collisions, $\delta_t$ is the time step and $f_{\alpha}^{eq}$ is
the truncated discrete form of the Maxwellian
\begin{equation}
f_{\alpha}^{eq}\equiv f_{\alpha}^{eq,M}(\rho,\mbox{\boldmath$u$})=
\omega_{\alpha}\left\{
1+\frac{\mbox{\boldmath$e$}_{\alpha} \cdotp \mbox{\boldmath$u$}}{RT}+
  \frac{\left( \mbox{\boldmath$e$}_{\alpha} \cdotp \mbox{\boldmath$u$} \right)^2}{2(RT)^2}-
  \frac{1}{2}\frac{\mbox{\boldmath$u$} \cdotp \mbox{\boldmath$u$}}{RT}
\right\},
\label{eq:trunceq}
\end{equation}
where $R$ is the gas constant, $T$ is the temperature and $w_{\alpha}$ is the weighting coefficients in the
Gauss-Hermite quadrature to represent the kinetic moment integrals of the distribution functions exactly~\cite{he97}.
For isothermal flows,
the factor $RT$ is related to the particle speed $c$ as $RT=1/3c^2$.
The term in Eq. (\ref{eq:axislbe})
\begin{equation}
S_{\alpha}=\frac{(e_{\alpha j}-u_j)(F_j+F_{ext,j})}{\rho RT}f_{\alpha}^{eq,M}(\rho,\mbox{\boldmath$u$})
\label{eq:sourcemp}
\end{equation}
represents the effect of internal and external forcing terms on the change in the distribution function.
The internal force term gives rise to surface tension and phase segregation effects which are given by
\begin{equation}
F_j=-\partial_j \psi + F_{s,j},
\label{eq:forceint}
\end{equation}
where the function $\psi=P-\rho RT$ is the non-ideal part of the equation of state given in Eq. (\ref{eq:axiseos}).
The first two terms on the RHS of Eq. (\ref{eq:axislbe}) corresponds to those presented by He \emph{et al.} (1998).
As mentioned above, the last two terms, $S_{\alpha}^{'}$ and $S_{\alpha}^{''}$, in this equation is
to be selected such that its behavior in the continuum limit would simulate the influence of the non-Cartesian-like
terms in Eqs. (\ref{eq:axiscont1}) and (\ref{eq:axismom1}) in a self-consistent way. Since
the zeroth kinetic moment of the term $f_{\alpha}^{eq,M}(\rho,0)$ is involved in the derivation of the macroscopic
mass conservation equation from the LBE, the source term $S_{\alpha}^{'}$  in Eq. (\ref{eq:axislbe}) is proposed to be equal to
$f_{\alpha}^{eq,M}(\rho,0)$ multiplied by an unknown $m^{'}$ and normalized by the density $\rho$. The other
source term $S_{\alpha}^{''}$ is proposed analogous to
the source term in Eq.(\ref{eq:sourcemp}). Thus, we propose
\begin{eqnarray}
S_{\alpha}^{'}&=&\frac{f_{\alpha}^{eq,M}(\rho,0)}{\rho}m^{'}\label{eq:sourcea1},\\
S_{\alpha}^{''}&=&\frac{(e_{\alpha j}-u_j)F_j^{''}}{\rho RT}f_{\alpha}^{eq,M}(\rho,\mbox{\boldmath$u$}).
\label{eq:sourcea2}
\end{eqnarray}
Here the unknowns, $m^{'}$ and $F_j^{''}$, in the above two equations can be determined through Chapman-Enskog
analysis as will be shown later. It must be stressed that all terms, including the collision term,
on the RHS are discretized by the application of the trapezoidal rule, since it has been argued that at least
a second-order treatment of the source terms is necessary for simulation of multiphase flow~\cite{he98,he99}.
The macroscopic fields are given by
\begin{eqnarray}
\rho&=&\sum_{\alpha} f_{\alpha},\\
\rho u_i&=&\sum_{\alpha} f_{\alpha} e_{\alpha i}.
\label{eq:amacrofields}
\end{eqnarray}
In this model, the order parameter is the density, $\rho$, which distinguishes the different phases in the flow.

Equation (\ref{eq:axislbe}) is implicit in time. To remove implicitness in this equation we introduce
a transformation following the procedure described by He and others~\cite{he98,he98a}, whereby
\begin{equation}
\bar{f}_{\alpha}=f_{\alpha}-\frac{1}{2}\Omega_{\alpha}-
\frac{1}{2}\left(S_{\alpha}+S_{\alpha}^{'}+S_{\alpha}^{''}\right)\delta_t
\label{eq:implicittr}
\end{equation}
in Eq. (\ref{eq:axislbe}), so that we obtain
\begin{equation}
\bar{f}_{\alpha}( \mbox{\boldmath$x$}+\mbox{\boldmath$e$}_{\alpha}\delta_t,t+\delta_t )-
\bar{f}_{\alpha}( \mbox{\boldmath$x$},t )=
\bar{\Omega}_{\alpha}|_{(x,t)}+\frac{\tau}{\tau+1/2}\left[ S_{\alpha}+S_{\alpha}^{'}+
S_{\alpha}^{''}\right]|_{(x,t)}\delta_t,
\label{eq:axislbee}
\end{equation}
where
\begin{equation}
\bar{\Omega}_{\alpha}=-\frac{\bar{f}_{\alpha}-f_{\alpha}^{eq}}{\tau+1/2}.
\label{eq:axisbgk}
\end{equation}
Thus, $\bar{f}_{\alpha}$ is the transformed distribution function that removes implicitness in the
proposed LBE, Eq. (\ref{eq:axislbe}), which describes the evolution of the $f_{\alpha}$ distribution function.
The following constraints on the equilibrium distribution and the various
source terms~\cite{luo00,guo02} are imposed from their definition:
\begin{equation}
\sum_{\alpha} f_{\alpha}^{eq}=\rho, \quad \sum_{\alpha} f_{\alpha}^{eq} e_{\alpha i}=\rho u_i, \quad
\sum_{\alpha} f_{\alpha}^{eq} e_{\alpha i}  e_{\alpha j}=\rho RT \delta_{ij}+\rho u_i u_j, \nonumber
\end{equation}
\begin{equation}
\sum_{\alpha} f_{\alpha}^{eq} e_{\alpha i}  e_{\alpha j} e_{\alpha k}=\rho (RT)^2
\left( u_i \delta_{jk}+u_j \delta_{ki}+u_k \delta_{ij}\right),
\end{equation}
\begin{equation}
\sum_{\alpha} S_{\alpha}=0, \quad \sum_{\alpha} S_{\alpha} e_{\alpha i} = F_i, \quad
\sum_{\alpha} S_{\alpha} e_{\alpha i} e_{\alpha j}=(F_i+F_{ext,i}) u_j+(F_j+F_{ext,j}) u_i,
\end{equation}
\begin{equation}
\sum_{\alpha} S_{\alpha}^{'}=m^{'}, \quad \sum_{\alpha} S_{\alpha}^{'} e_{\alpha i} = 0, \quad
\sum_{\alpha} S_{\alpha}^{'} e_{\alpha i} e_{\alpha j}=m^{'}RT\delta_{ij},
\end{equation}
\begin{equation}
\sum_{\alpha} S_{\alpha}^{''}=0, \quad \sum_{\alpha} S_{\alpha}^{''} e_{\alpha i} = F_i^{''}, \quad
\sum_{\alpha} S_{\alpha}^{''} e_{\alpha i} e_{\alpha j}=(F_i^{''} u_j+F_j^{''} u_i).
\end{equation}
Then the following relationships are obtained between the transformed distribution function and the macroscopic
fields, which also include the curvature effects resulting from axial symmetry:
\begin{eqnarray}
\rho&=&\sum_{\alpha} \bar{f}_{\alpha}+\frac{1}{2}m^{'}\delta_t \label{eq:dens1},\\
\rho u_i&=&\sum_{\alpha} \bar{f}_{\alpha} e_{\alpha i}+\frac{1}{2}(F_i+F_{ext,i}+F_i^{''})\delta_t.
\label{eq:densvel1}
\end{eqnarray}

Now, to establish the unknowns $m^{'}$ and $F_j^{''}$ in the above formulation, the Chapman-Enskog multiscale
analysis is performed~\cite{chapman}. Introducing the expansions~\cite{he97a}
\begin{eqnarray}
\bar{f}_{\alpha}( \mbox{\boldmath$x$}+\mbox{\boldmath$e$}_{\alpha}\delta_t,t+\delta_t )&=&
\sum_{\alpha=0}^{\infty} D_{t_n}\bar{f}_{\alpha}(\mbox{\boldmath$x$},t),\\
D_{t_n} &\equiv& \partial_{t_n}+e_{\alpha k} \partial_k,\\
f_{\alpha}&=&\sum_{\alpha=0}^{\infty}\epsilon^n f_{\alpha}^{(n)},\\
\partial_t&=&\sum_{\alpha=0}^{\infty}\epsilon^n \partial_{t_n},
\end{eqnarray}
where $\epsilon=\delta_t$ in Eq. (\ref{eq:axislbee}) and using Eq. (\ref{eq:implicittr}) to transform $\bar{f}_\alpha$ back to
$f_\alpha$, the following equations are obtained in
the consecutive order of the parameter $\epsilon$:
\begin{eqnarray}
O(\epsilon^0): f_{\alpha}^{(0)}&=&f_{\alpha}^{eq}\label{eq:order0},\\
O(\epsilon^1): D_{t_0} f_{\alpha}^{(0)}&=&-\frac{1}{\tau} f_{\alpha}^{(1)}+S_{\alpha}+
S_{\alpha}^{'}+S_{\alpha}^{''}\label{eq:order1},\\
O(\epsilon^2): \partial_{t_1} f_{\alpha}^{(0)}+ D_{t_0} f_{\alpha}^{(1)}&=&-\frac{1}{\tau} f_{\alpha}^{(2)}.
\label{eq:order2}
\end{eqnarray}
Now, invoking the Chapman-Enskog ansatz
\begin{equation}
\sum_{\alpha}
\left( \begin{array}{c}
1 \\
e_{\alpha i}
\end{array} \right)f_{\alpha}^{(0)}=
\left( \begin{array}{c}
\rho \\
\rho u_i
\end{array} \right),
\sum_{\alpha}
\left( \begin{array}{c}
1 \\
e_{\alpha i}
\end{array} \right)f_{\alpha}^{(n)}=
\left( \begin{array}{c}
0 \\
0
\end{array} \right),n \geq 1
\end{equation}
and performing $\sum_{\alpha}(\cdotp)$ on Eqs. (\ref{eq:order1}) and (\ref{eq:order2}), we obtain
\begin{eqnarray}
\partial_{t_0} \rho + \partial_k (\rho u_k)&=& m^{'} \label{eq:axisc1},\\
\partial_{t_1} \rho                        &=& 0,\label{eq:axisc2}
\end{eqnarray}
respectively. Combining the first- and second- order results given by Eqs. (\ref{eq:axisc1}) and (\ref{eq:axisc2})
and considering $\partial_t=\partial_{t_0}+\epsilon \partial_{t_1}$, we get
\begin{equation}
\partial_t \rho + \partial_k (\rho u_k)= m^{'} \label{eq:axisc}.\\
\end{equation}
Comparing this equation and Eq. (\ref{eq:axiscont1}), the unknown $m^{'}$ is
obtained as
\begin{equation}
m^{'}=-\frac{\rho u_y}{y}. \label{eq:mvalue}\\
\end{equation}
This is the axisymmetric contribution to the Cartesian form of the equation for the order parameter,
i.e., density characterizing the different phases of the flow. Taking the first kinetic moment,
$\sum_{\alpha}e_{\alpha i}(\cdotp)$, of  Eqs. (\ref{eq:order1}) and (\ref{eq:order2}), respectively, we get
\begin{eqnarray}
\partial_{t_0} (\rho u_i) + \partial_k (\rho u_i u_k)&=& -\partial_i (\rho RT)+F_i+F_{ext,i}+F_i^{''}, \label{eq:axism1}\\
\partial_{t_1} (\rho u_i) + \partial_k \Pi_{ij}^{(1)}&=& 0, \label{eq:axism2}
\end{eqnarray}
where
\begin{equation}
\Pi_{ij}^{(1)}=\sum_{\alpha} f_{\alpha}^{(1)}e_{\alpha i} e_{\alpha j}.
\label{eq:visct1}
\end{equation}
Employing the expression
for $f_{\alpha}^{(1)}$ from Eq. (\ref{eq:order1}) in Eq. (\ref{eq:visct1}), together with the summational
constraints given above, and neglecting terms of the order $O(Ma^3)$ or higher, we get
\begin{equation}
\Pi_{ij}^{(1)}=-\tau RT \rho (\partial_j u_i+\partial_i u_j).
\label{eq:visct2}
\end{equation}
Equation (\ref{eq:axism2}) then simplifies to
\begin{equation}
\partial_{t_1} (\rho u_i) = \partial_j \left( \tau RT \rho (\partial_j u_i+\partial_i u_j)  \right) \label{eq:axism2n}.
\end{equation}
Combining Eqs. (\ref{eq:axism1}) and (\ref{eq:axism2n}), we get
\begin{equation}
\partial_t (\rho u_i) + \partial_k (\rho u_i u_k)= -\partial_i (\rho RT)+F_i+F_{ext,i}+F_i^{''}+
\partial_j \left( \tau \delta_t RT \rho (\partial_j u_i+\partial_i u_j)  \right),
\end{equation}
or substituting for $F_i$ from Eq. (\ref{eq:forceint}), we obtain
\begin{equation}
\partial_t (\rho u_i) + \partial_k (\rho u_i u_k)= -\partial_i P +F_{s,i}+F_{ext,i}+F_i^{''}+
\partial_j \left( \tau \delta_t RT \rho (\partial_j u_i+\partial_i u_j)  \right).
\label{eq:axism3}
\end{equation}
Using Eqs. (\ref{eq:axisc}) and (\ref{eq:axism3}), this can be simplified to
\begin{eqnarray}
\rho \left( \partial_t u_i + u_k \partial_k u_i \right)-\frac{\rho u_i u_y}{y}&=&
-\partial_i P +F_{s,i}+F_{ext,i}+F_i^{''}+\nonumber\\
& & \partial_j \left( \tau \delta_t RT \rho (\partial_j u_i+\partial_i u_j)  \right).
\label{eq:axism4}
\end{eqnarray}
Comparing  Eqs. (\ref{eq:axismom1}) and (\ref{eq:axism4}), we obtain the other unknown
$F_i^{''}$ where
\begin{equation}
F_i^{''}=F_{ax,i}-\frac{\rho u_i u_y}{y}=\frac{\mu}{y}\left[ \partial_y u_i + \partial_i u_y \right]+
\kappa \rho \partial_i \left(\frac{1}{y}\partial_y \rho \right)-
\frac{\rho u_i u_y}{y}.
\label{eq:fvalue}
\end{equation}
This is the axisymmetric contribution to the Cartesian form of the equation for the momentum, where
the first, second and the third terms on the RHS correspond to the viscous, surface tension and inertial force contributions,
respectively. The dynamic viscosity is related to the relaxation
time for collisions by $\mu=\rho \tau \delta_t RT = \rho \lambda c_s^2$, where $c_s^2=1/3c^2$. The set of equations
corresponding to the axisymmetric LBE multiphase flow model is given by Eqs. (\ref{eq:axislbee}) and (\ref{eq:axisbgk}) together with
Eqs. (\ref{eq:sourcemp}), (\ref{eq:sourcea1}) and (\ref{eq:sourcea2}), (\ref{eq:dens1}) and (\ref{eq:densvel1}), and
(\ref{eq:mvalue}) and (\ref{eq:fvalue}). In general, this multiphase model and that proposed by He and others~\cite{he98}
face difficulties for fluids far from the critical point and/or in the presence of external forces. This
difficulty is related to the calculation of the intermolecular force in Eq.(\ref{eq:forceint}), involving
the computation of $\partial_j\psi$ which can become quite large across interfaces. Unless this term
is accurately computed, the model may become unstable because of numerical errors~\cite{he99a,he04}. Hence, an
improved treatment of this term is necessary. This will now be described.

\section{\label{sec:axismodelc}Axisymmetric LBE Multiphase Flow Model with Reduced Compressibility Effects}
He and co-workers~\cite{he99} have proposed that through a suitable transformation of the
distribution function, $f_{\alpha}$, which involves invoking the incompressibility condition of the fluid,
and employing a new distribution function for capturing the interface, the difficulty with handling the
intermolecular force term, $\partial_j\psi$, can be reduced. We apply this idea to the axisymmetric
model developed in the previous section. We replace
the distribution function $f_{\alpha}$ by another distribution function $g_{\alpha}$ through
the transformation~\cite{he99}
\begin{equation}
g_{\alpha}=f_{\alpha}RT+\psi(\rho)\frac{f_{\alpha}^{eq,M}(\rho,0)}{\rho}.
\label{eq:transg}
\end{equation}
The effect of this transformation will be discussed in greater detail below. By considering the fluid to be incompressible, i.e.
\begin{equation}
\frac{d}{dt}\psi(\rho)=\left( \partial_t+u_k \partial_k \right)\psi (\rho)=0,
\end{equation}
and using the transformation Eqs. (\ref{eq:transg}) and (\ref{eq:implicittr}), Eq. (\ref{eq:axislbee}) is replaced by
\begin{equation}
\bar{g}_{\alpha}( \mbox{\boldmath$x$}+\mbox{\boldmath$e$}_{\alpha}\delta_t,t+\delta_t )-
\bar{g}_{\alpha}( \mbox{\boldmath$x$},t )=
\bar{\Omega}_{g\alpha}|_{(x,t)}+\frac{\tau}{\tau+1/2}\left[ S_{g\alpha}+S_{g\alpha}^{'}+
S_{g\alpha}^{''}\right]|_{(x,t)}\delta_t,
\label{eq:axislbeg}
\end{equation}
where
\begin{equation}
\bar{\Omega}_{g\alpha}=-\frac{\bar{g}_{\alpha}-g_{\alpha}^{eq}}{\tau+1/2},
\label{eq:axisbgkg}
\end{equation}
and
\begin{equation}
g_{\alpha}^{eq}=f_{\alpha}^{eq}RT+\psi(\rho)\frac{f_{\alpha}^{eq,M}(\rho,0)}{\rho}.
\end{equation}
The corresponding source terms become
\begin{eqnarray}
S_{g\alpha}&=&(e_{\alpha j}-u_j)\times \nonumber\\
	   & & \left[
	       (F_j+F_{ext,j})\frac{f_{\alpha}^{eq,M}(\rho,\mbox{\boldmath$u$})}{\rho}-
       \left(
       \frac{f_{\alpha}^{eq,M}(\rho,\mbox{\boldmath$u$})}{\rho}-\frac{f_{\alpha}^{eq,M}(\rho,0)}{\rho}
       \right)\partial_j \psi(\rho)
       \right],
\label{eq:srcrefined}
\end{eqnarray}
\begin{equation}
S_{g\alpha}^{'}=S_{\alpha}^{'}RT=\frac{f_{\alpha}^{eq,M}(\rho,0)}{\rho}\left( -\frac{\rho u_y}{y} \right)RT,
\end{equation}
\begin{equation}
S_{g\alpha}^{''}=S_{\alpha}^{''}RT=(e_j-u_j)F_j^{''}\frac{f_{\alpha}^{eq,M}(\rho,\mbox{\boldmath$u$})}{\rho}.
\label{eq:sourceg2}
\end{equation}
The term $\partial_j \psi$ in Eq. (\ref{eq:srcrefined}) is multiplied by the factor
$\left( f_{\alpha}^{eq,M}(\rho,\mbox{\boldmath$u$})/\rho-f_{\alpha}^{eq,M}(\rho,0)/\rho \right)$. This factor,
from the definition of the equilibrium distribution function, $f_{\alpha}^{eq}$, in Eq. (\ref{eq:trunceq})
is proportional to the Mach number and thus becomes smaller in the incompressible limit. Hence, it alleviates
the difficulties associated with the calculation of the $\partial_j \psi$, a major source of numerical
instability with the original model~\cite{he98}.
Thus, Eqs. (\ref{eq:axislbeg})-(\ref{eq:sourceg2}) are found to be numerically more stable compared to
Eq. (\ref{eq:axislbee}) supplemented with Eqs. (\ref{eq:sourcemp}),(\ref{eq:sourcea1}) and (\ref{eq:sourcea2}).
In this new framework, we still need to introduce
an order parameter to capture interfaces. Here, we employ a function, $\phi$, referred to henceforth
as the index function, in place of the density, as the order parameter to distinguish the phases in the flow.

The evolution equation of the distribution function whose emergent dynamics govern the index function has
to be able to maintain phase segregation and mass conservation. To do this, we employ
Eq. (\ref{eq:axislbee}) together with Eqs. (\ref{eq:sourcemp}),(\ref{eq:sourcea1}) and (\ref{eq:sourcea2}) by keeping
the term involving $\partial_j \psi$ and $m^{'}$, while the rest of the terms may be dropped as they
play no role in mass conservation. In addition, the density is replaced by the index function in these
equations. Hence, the evolution of the distribution function for the index function is given by
\begin{equation}
\bar{f}_{\alpha}( \mbox{\boldmath$x$}+\mbox{\boldmath$e$}_{\alpha}\delta_t,t+\delta_t )-
\bar{f}_{\alpha}( \mbox{\boldmath$x$},t )=
\bar{\Omega}_{f\alpha}|_{(x,t)}+\frac{\tau}{\tau+1/2}\left[ S_{f\alpha}+
S_{f\alpha}^{'}\right]|_{(x,t)}\delta_t,
\label{eq:axislbef}
\end{equation}
where the collision and the source terms are given by
\begin{equation}
\bar{\Omega}_{f\alpha}=-\frac{\bar{f}_{\alpha}-\frac{\phi}{\rho}f_{\alpha}^{eq}}{\tau+1/2},
\label{eq:axisbgkf}
\end{equation}
\begin{equation}
S_{f\alpha}=\frac{(e_j-u_j)(-\partial_j \psi(\phi))}{\rho RT}f_{\alpha}^{eq,M}(\rho,\mbox{\boldmath$u$}),
\label{eq:srciref}
\end{equation}
\begin{equation}
S_{f\alpha}^{'}=\frac{\phi}{\rho}S_{\alpha}^{'}=
\frac{f_{\alpha}^{eq,M}(\rho,0)}{\rho}\left( -\frac{\phi u_y}{y} \right).
\end{equation}
The hydrodynamic variables such as pressure and fluid velocity can be
obtained by taking appropriate kinetic moments of the distribution function $g_{\alpha}$, i.e.
\begin{eqnarray}
P&=&\sum_{\alpha} \bar{g}_{\alpha}-\frac{1}{2}u_j\partial_j \psi(\rho)+\frac{1}{2}m^{'}RT\delta_t,\label{eq:axpres}\\
\rho RT u_i &=& \sum_{\alpha} \bar{g}_{\alpha} e_{\alpha i}+\frac{1}{2}\left( F_{s,i}+F_{ext,i} \right)\delta_t+
\frac{1}{2} F_i^{''}\delta_t.
\end{eqnarray}
This follows from the definition of $\bar{g}_{\alpha}$ given in Eq. (\ref{eq:transg}) and also includes curvature effects.
The index function is obtained from the distribution function $\bar{f}_{\alpha}$ by taking the zeroth
kinetic moment, i.e.
\begin{equation}
\phi=\sum_{\alpha}\bar{f}_{\alpha}+\frac{1}{2}\frac{\phi}{\rho}m^{'}\delta_t.
\end{equation}
The terms $m^{'}$ and $F_i^{''}$ are given in Eqs. (\ref{eq:mvalue}) and (\ref{eq:fvalue}),
respectively. The density is obtained from the index function through linear interpolation, i.e.
\begin{equation}
\rho(\phi)= \rho_L+\frac{\phi-\phi_L}{\phi_H-\phi_L}(\rho_H-\rho_L),
\label{eq:rintp}
\end{equation}
where $\rho_L$ and $\rho_H$ are the densities of the light and heavy fluids, respectively, and $\phi_L$ and
$\phi_H$ refer to the minimum and maximum values of the index function, respectively. These limits of the index
function are determined from Maxwell's equal area construction~\cite{rowlinson} applied to the function
$\psi(\phi)+\phi RT$.

Thus, the axisymmetric LBE multiphase flow model with reduced compressibility effects corresponds 
to Eqs. (\ref{eq:axislbeg})-(\ref{eq:rintp}).
The relaxation time for collisions is related to the viscosity of the fluid using the same
expression as derived in the previous section. If the kinematic viscosity of the light fluid, $\nu_L$,
is different from that of the heavy fluid, $\nu_H$, its value at any point in the fluid is obtained
from the index function through linear interpolation, i.e.
\begin{equation}
\nu(\phi)= \nu+\frac{\phi-\phi_L}{\phi_H-\phi_L}(\nu-\nu_L).
\label{eq:vintp}
\end{equation}

It may be seen that the model requires the calculation of spatial gradients in Eqs. (\ref{eq:srcrefined}) and (\ref{eq:srciref})
and of the Laplacian in Eq. (\ref{eq:forcemp}).
Since maintaining
accuracy as well as isotropy is important for the surface tension terms, they are calculated
by employing a fourth-order finite-difference scheme for the gradient and a second-order scheme
for the Laplacian, given respectively by
\begin{equation}
\partial_i \varpi=\frac{1}{36\delta_x}\sum_{\alpha=1}^{8}\left[
8\varpi(\mbox{\boldmath$x$}+\mbox{\boldmath$e$}_{\alpha i}\delta_t)-
\varpi(\mbox{\boldmath$x$}+2\mbox{\boldmath$e$}_{\alpha i}\delta_t)
\right] \left( \frac{e_{\alpha i}}{c} \right)+O(\delta_t^4),
\end{equation}
and
\begin{equation}
\nabla^2 \varpi \equiv \partial_i \partial_i \varpi = \frac{1}{3\delta_x^2}
\sum_{\alpha=1}^8\left[
\varpi(\mbox{\boldmath$x$}+\mbox{\boldmath$e$}_{\alpha i}\delta_t)-
\varpi(\mbox{\boldmath$x$})
\right]+O(\delta_x^2),
\end{equation}
for any function $\varpi$. Notice that these discretizations are both based on the lattice based stencil, instead of the standard
stencil based on the coordinate directions.
In addition, in the application of this model, the implementation of boundary
conditions plays an important role. In particular, along the axisymmetric line, i.e. $y=0$, specular
reflection boundary conditions are employed for the distribution functions. For the two-dimensional,
nine velocity (D2Q9) model shown in the inset of Fig.~\ref{fig:schemaxis}, we set
$\bar{f}_2=\bar{f}_4$, $\bar{f}_5=\bar{f}_8$,
$\bar{f}_6=\bar{f}_7$ and $\bar{g}_2=\bar{g}_4$, $\bar{g}_5=\bar{g}_8$ and $\bar{g}_6=\bar{g}_7$ for
the distribution functions after the streaming step. For macroscopic conditions, along this line,
$u_y=\partial_y(\cdotp)=0$, through which the singular source terms of type $1/y(\cdotp)$ in the model
can be appropriately treated. On the other hand, boundary conditions along the other lines are similar to
those for the standard LBE.

\section{\label{sec:results}Results and Discussion}
\begin{figure}
\begin{center}
\includegraphics{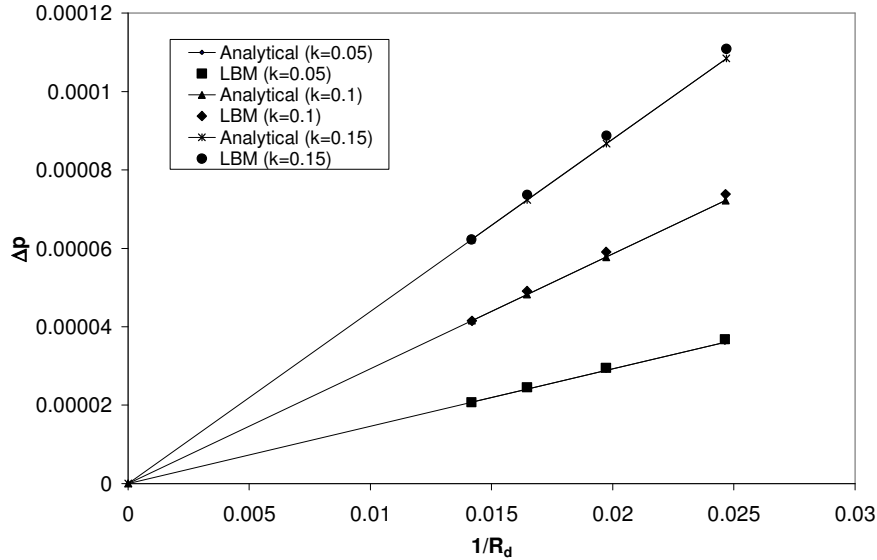}
\caption{\label{fig:laplaceyoung1}Pressure difference across axisymmetric drops as a function of radius for different 
         values of the surface tension  parameter $\kappa$; Comparison of computed results using the axisymmetric LBE model versus
	 theoretical prediction based on the Laplace-Young relation. Quantities are in lattice units.}
\end{center}
\end{figure}

In the rest of this paper, unless otherwise specified, the results are presented in lattice units, i.e. the velocities
are scaled by the particle velocity $c$, the distance by the minimum lattice spacing $\delta_x$ and time by $c/\delta_x$.
All other quantities are scaled as appropriate combinations of these basic units.
First, the axisymmetric LBE multiphase flow models are applied to verify the well-known
Laplace-Young relation for an axisymmetric drop. According to this relation,
$\Delta P=2 \sigma/R_d$, where $\Delta P$ is the difference
between the pressure inside and outside of a drop, $\sigma$ is the surface tension and $R_d$ is the drop radius.
For different choices of the surface tension parameter, $\kappa$, the surface tension values are obtained
from Eq. (\ref{eq:sigmakappa}) by the replacing density in Eq. (\ref{eq:surfr})  and (\ref{eq:surfz})
by the index function. To obtain the normal gradient used in Eq. (\ref{eq:sigmakappa}), a physical
configuration consisting of a liquid and a gas layer is set up. Once equilibrium is reached, the density
gradient may be computed and hence the surface tension. Having obtained the relationship between the surface
tension $\sigma$, and the parameter $\kappa$, axisymmetric drops of four different radii, $R_d=40, 50, 60$
and $70$, are set up in a domain discretized by $201\times 101$ lattice sites. Periodic
boundaries are considered in the $x$ direction and an open boundary condition is considered along the boundary
that is parallel to the axisymmetric boundary. By considering three different values of
$\kappa$, $0.05, 1.0$ and $0.15$, the pressure difference across the drops is determined.
Figure~\ref{fig:laplaceyoung1} shows a comparison of the pressure difference across the 
interface of the drops computed using the axisymmetric
model developed in Section \ref{sec:axismodelc} and that predicted by
the Laplace-Young relation. It is found that the computed results are in good agreement with the theoretical values, with
a maximum relative error of about $3\%$.

Another important test problem is that of an oscillating axisymmetric drop immersed
in a gas. Since current versions of the LBE simulate a relatively viscous fluid, it is
appropriate to compare the oscillation frequency with that of Miller and Scriven (1968)~\nocite{miller68}.
In contrast to earlier analytical solutions on
drop oscillations, this work considers viscous dissipation effects in the boundary layer at the interface.
According to~\cite{miller68}, the frequency for the $n^{th}$ mode of oscillation for a drop is given by
\begin{equation}
\omega_{n}=\omega_{n}^{*}-\frac{1}{2} \alpha \omega_{n}^{*\frac{1}{2}}+\frac{1}{4}\alpha^2,
\label{eq:msperiod}
\end{equation}
where $\omega_{n}$ is the angular response frequency, and $\omega_{n}^{*}$ is Lamb's natural resonance
frequency expressed as~\cite{lamb}
\begin{equation}
\left(\omega_{n}^{*}\right)^{2}=
\frac{n(n+1)(n-1)(n+2)}{R_d^3 \left[ n\rho_{g}+(n+1)\rho_{l} \right]} \sigma.
\end{equation}
$R_d$ is the equilibrium radius of the drop, $\sigma$ is the interfacial surface tension, and $\rho_{l}$ and
$\rho_{g}$ are the densities of the two fluids. The parameter $\alpha$ is given by
\begin{equation}
\alpha =
\frac{(2n+1)^2 (\mu_{l} \mu_{g} \rho_{l} \rho_{g})^{\frac{1}{2}}}
{2^{\frac{1}{2}} R_d \left[ n\rho_{g}+(n+1)\rho_{l} \right]
\left[ (\mu_{l} \rho_{l})^{\frac{1}{2}}+ (\mu_{g} \rho_{g})^{\frac{1}{2}} \right]},
\end{equation}
where $\mu_{l}$ and $\mu_{g}$ are the dynamic viscosity of the two liquids. The subscripts $g$ and $l$
refer to the ambient gas and liquid phases, respectively. We consider the second mode of oscillation and
analytical expressions for the time period are presented in Eq. (\ref{eq:msperiod}).

\begin{figure}
\begin{center}
\includegraphics{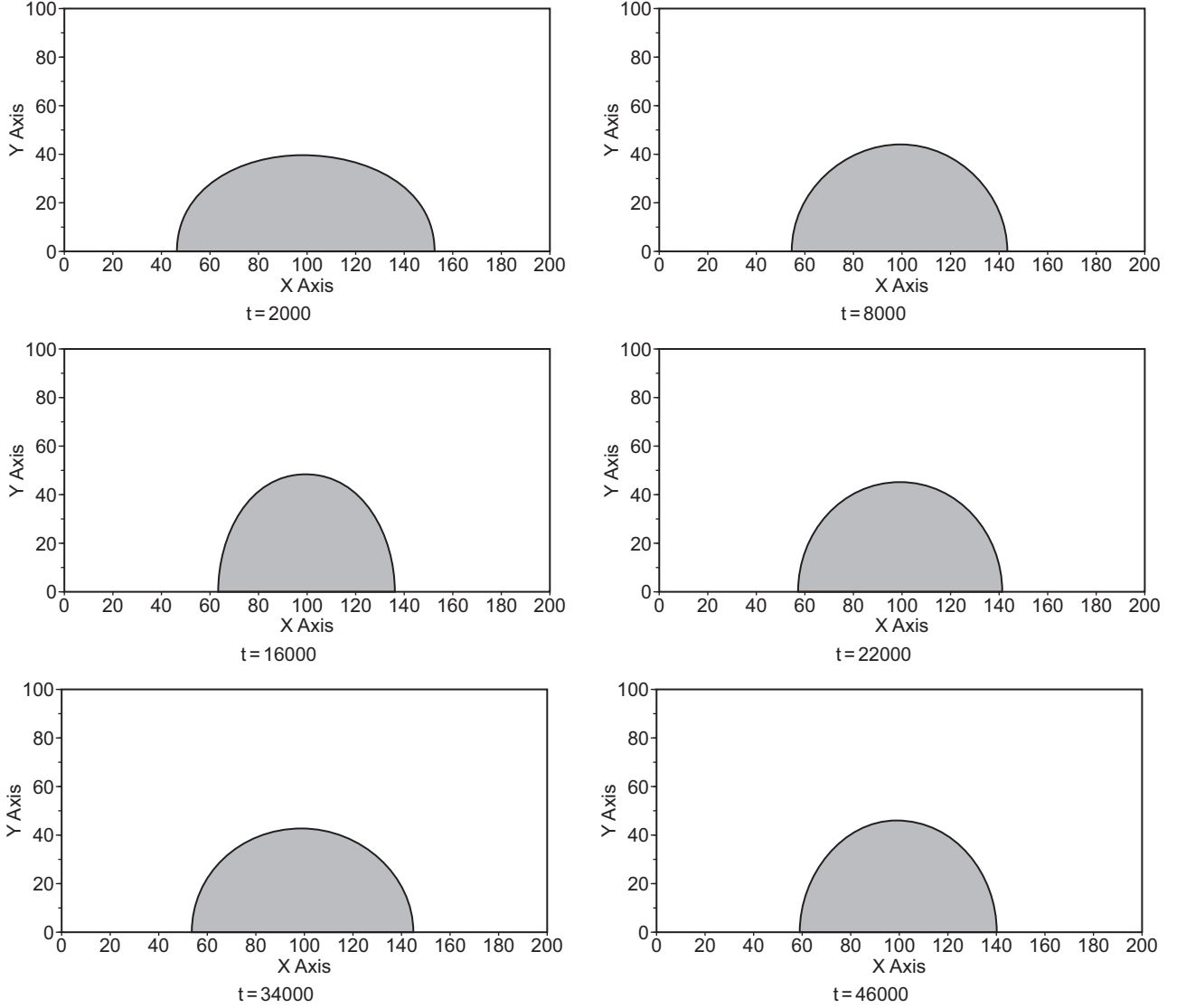}
\caption{\label{fig:oscchem}Configurations of an oscillating drop as a function of time; 
         $R_{min}=40$, $R_{max}=55$,
	 $\rho_g=0.1$, $\rho_l=0.4$, $\nu_l=\nu_g=1.6667\times 10^{-2}$. Quantities are in lattice units.}
\end{center}
\end{figure}

\begin{figure}
\begin{center}
\includegraphics{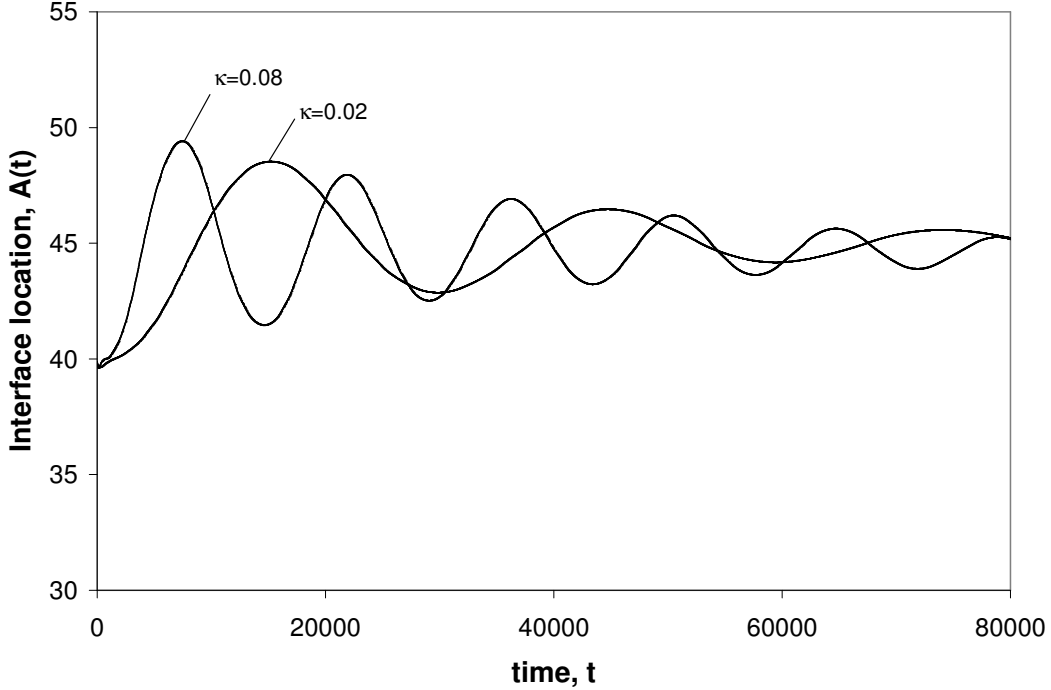}
\caption{\label{fig:dropint1}Interface location of an oscillating drop as a function of time for 
         two values of the surface tension
	 parameter $\kappa$; $R_{min}=40$, $R_{max}=55$, $\rho_g=0.1$, $\rho_l=0.4$, $\nu_l=\nu_g=1.6667\times 10^{-2}$. Quantities are in lattice units.}
\end{center}
\end{figure}

The initial computational setup consists of a prolate spheroid of minimum ($R_{min}$) and maximum ($R_{max}$)
radii of $40$ and $55$,
respectively, placed in the center of the domain discretized by $201 \times 101$ lattice sites. We consider
the surface tension parameters: $\kappa=0.2$, and the density of the gas and the drop to be $\rho_g=0.1$ and $\rho_l=0.4$,
respectively. The kinematic viscosity of both the gas and the drop are considered to be the same and given by
$\nu_g=\nu_l=1.6667\times 10^{-2}$. Figure~\ref{fig:oscchem} shows the configurations of an oscillating drop at different times
computed using the standard axisymmetric model with these conditions.
The drop changes from a
prolate shape at $t=2000$ to oblate shape at $t=16000$. Such shape changes continue till the drop reaches its
equilibrium spherical shape.
Figure \ref{fig:dropint1} shows the temporal evolution of the interface locations of the oscillating drop with the conditions above
for two different surface tension parameter: $\kappa=0.02$ and $0.08$.
It is expected that increasing the surface tension
will reduce the time period of oscillations. The computed ($T_{LBE}$) and analytical ($T_{anal}$) time periods, where 
$T_{anal}=2\pi/\omega_2$, when
$\kappa=0.02$ are $29483$ and $29448$ respectively. As $\kappa$ is increased to $0.08$, $T_{LBE}$ and $T_{anal}$ become
$14388$ and $14313$ respectively. It may be seen that the computed and analytical values agree well, the
difference being less than $1\%$. Also, the time period decreases as $\kappa$ is increased, which is consistent with
expectations.

\begin{figure}
\begin{center}
\includegraphics{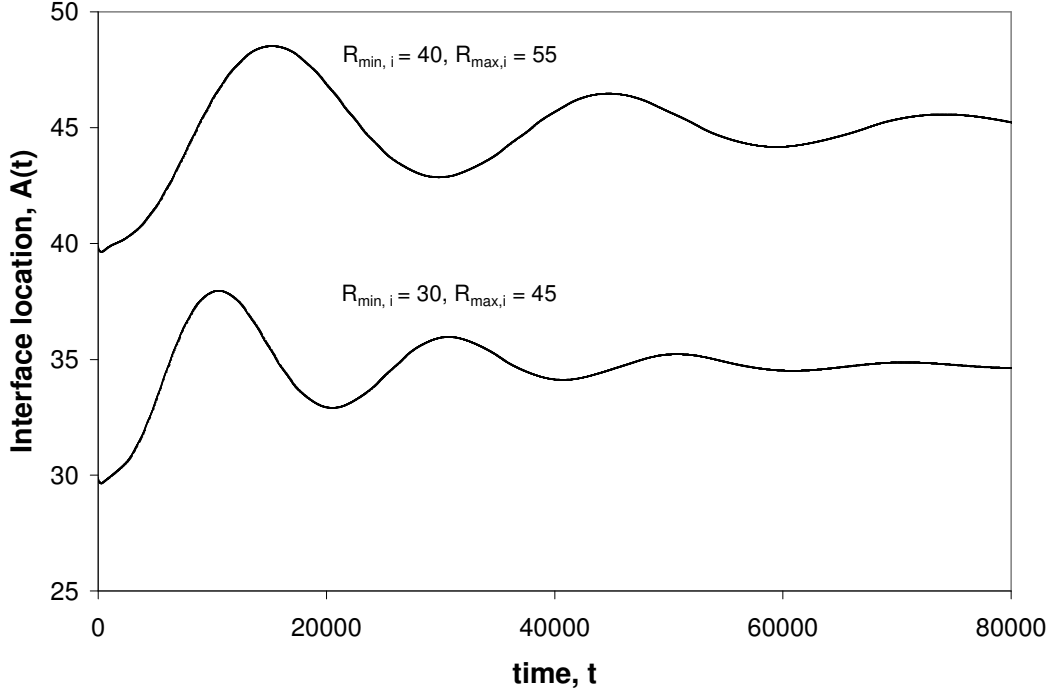}
\caption{\label{fig:dropint2}Interface location of an oscillating drop as a function of time for two drop sizes;
	 $\rho_g=0.1$, $\rho_l=0.4$, $\nu_l=\nu_g=1.6667\times 10^{-2}$, $\kappa=0.02$. Quantities are in lattice units.}
\end{center}
\end{figure}

\begin{figure}
\begin{center}
\includegraphics{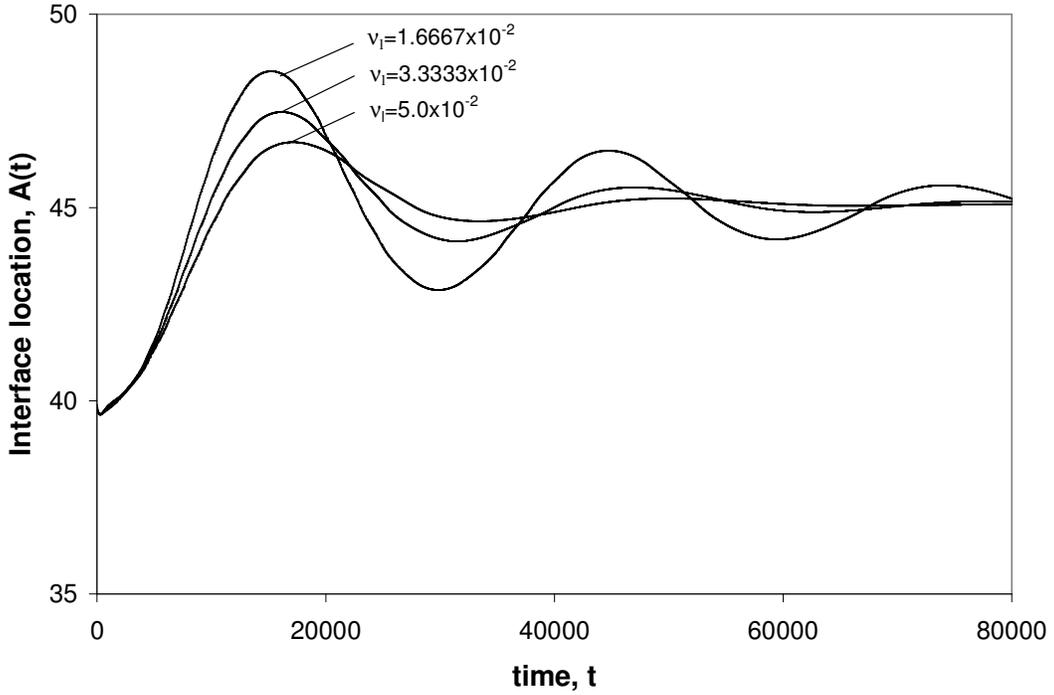}
\caption{\label{fig:dropint3}Interface location of an oscillating drop as a function of time for different kinematic viscosities $\nu_l$;
         $R_{min}=40$, $R_{max}=55$, $\rho_g=0.1$, $\rho_l=0.4$, $\kappa=0.02$. Quantities are in lattice units.}
\end{center}
\end{figure}

Consider next the effect of changing the drop size on the time period of oscillations.
Figure \ref{fig:dropint2} shows the interface locations of an oscillating drop as a function of time for the following two initial sizes:
$R_{min}=30$ and $R_{max}=45$; $R_{min}=40$, $R_{max}=55$. Reducing the drop size reduces its time period.
The computed time period of the larger drop is equal to $29483$, while that for the smaller drop is $20118$.
Comparison of the computed time periods with the analytical solution shows that they agree within $1\%$ for these cases.
Next, consider three different kinematic viscosities of the liquid: $\nu_l=1.6667\times 10^{-2}, 3.3333\times 10^{-2}$ and
$5.0\times 10^{-2}$.
Figure \ref{fig:dropint3} shows the effect of drop viscosity on the temporal evolution of the interface locations of the drop.
It is found that as the kinematic viscosity is increased the time period increases moderately which
is consistent with the analytical solution. The computed time periods at these
viscosities are $29483$, $31030$ and $32925$, while the analytical values are $29448$, $30597$ and $31318$, respectively, with
a maximum error within $5.1\%$.

\begin{figure}
\begin{center}
\includegraphics[height=6.50in,clip=]{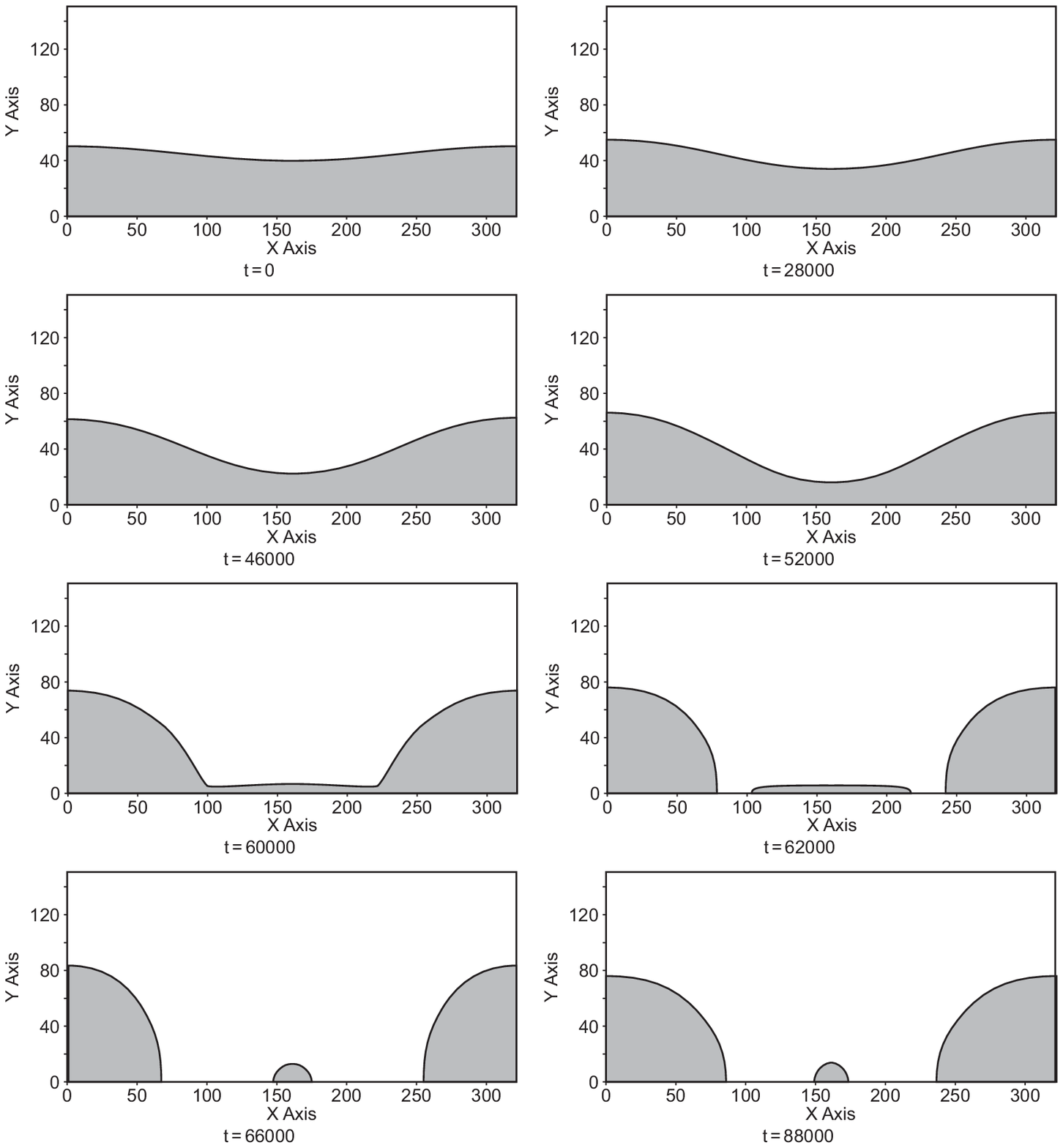}
\caption{\label{fig:rayleigh1}Configurations of a cylindrical liquid column at different times undergoing Rayleigh 
	 breakup and satellite droplet formation;
	 $k^{*}=0.88$, $\rho_g=0.1$, $\rho_l=0.4$, $\nu_l=\nu_g=6.6667\times 10^{-2}$. Quantities are in lattice units.}
\end{center}
\end{figure}

The third test problem considered here is that of the break-up of a cylindrical liquid column into drops, a fascinating
problem of long standing theoretical and practical interest. In a seminal work,
Rayleigh (1878)~\nocite{rayleigh78} showed through a linear stability analysis of an inviscid column of
cylindrical liquid of radius $R_c$ that the column will be unstable if the axisymmetric wavelength of any
disturbance $\lambda_d$ is longer than its circumference, i.e. the wave number
$k^{*}=2\pi R_c/\lambda_d$ should be less than one. Later, the theoretical analysis was extended to more
realistic conditions by including viscosity. In the last three decades, several experimental and numerical
investigations have also been performed. To evaluate the axisymmetric LBE model, we study the Rayleigh capillary
instability for different wavenumbers.  Initial studies carried out with $k^{*}>1$ showed that the liquid does not
break-up. We will now present results of cases with break-up.
Consider a cylindrical liquid column of radius $R_c=45$
subject to an axisymmetric co-sinusoidal wavelength $\lambda_d=320$, i.e. $k^{*}=0.88$.
To simulate the dynamics of instability for this wavenumber, we consider a domain discretized by $321\times 151$ lattice
sites with $\rho_g=0.1$, $\rho_l=0.4$, $\nu_g=\nu_l=6.6667\times 10^{-2}$ and $\kappa=0.1$. Since $k^{*}<1$, it is
expected that the liquid column would eventually breakup. Figure \ref{fig:rayleigh1} shows the configurations of the liquid
column at different times. As time progresses, the imposed interfacial disturbances on the
liquid column grow. At $t=28000, 46000$ $52000$, the cross-section of the column becomes
progressively thinner in the center, and by mass conservation, the ends becomes larger. At $t=60000$, notice
that a bead-type structure is formed at the ends and with a thin ligament between them. Such a structure has been
observed in experiments~\cite{eggers97} and in other numerical simulations~\cite{ashgriz95}. Eventually, the column
breaks up forming a thin ligament in the middle, which then becomes a satellite droplet.

\begin{figure}
\begin{center}
\includegraphics{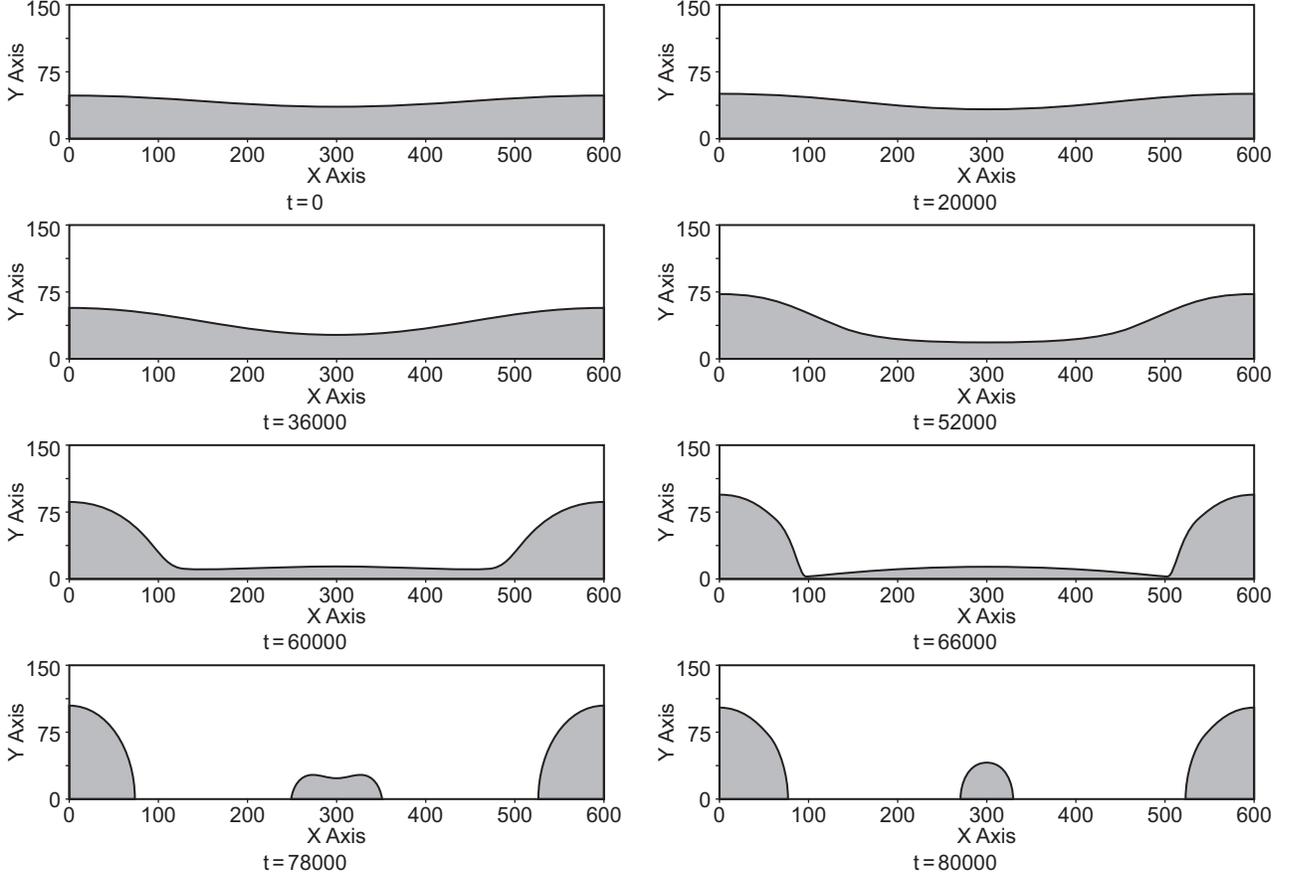}
\caption{\label{fig:rayleigh2}Configurations of a cylindrical liquid column at different times undergoing Rayleigh 
         breakup and satellite droplet formation;
	 $k^{*}=0.47$, $\rho_g=0.1$, $\rho_l=0.4$, $\nu_l=\nu_g=6.6667\times 10^{-2}$. Quantities are in lattice units.}
\end{center}
\end{figure}

\begin{figure}
\begin{center}
\includegraphics{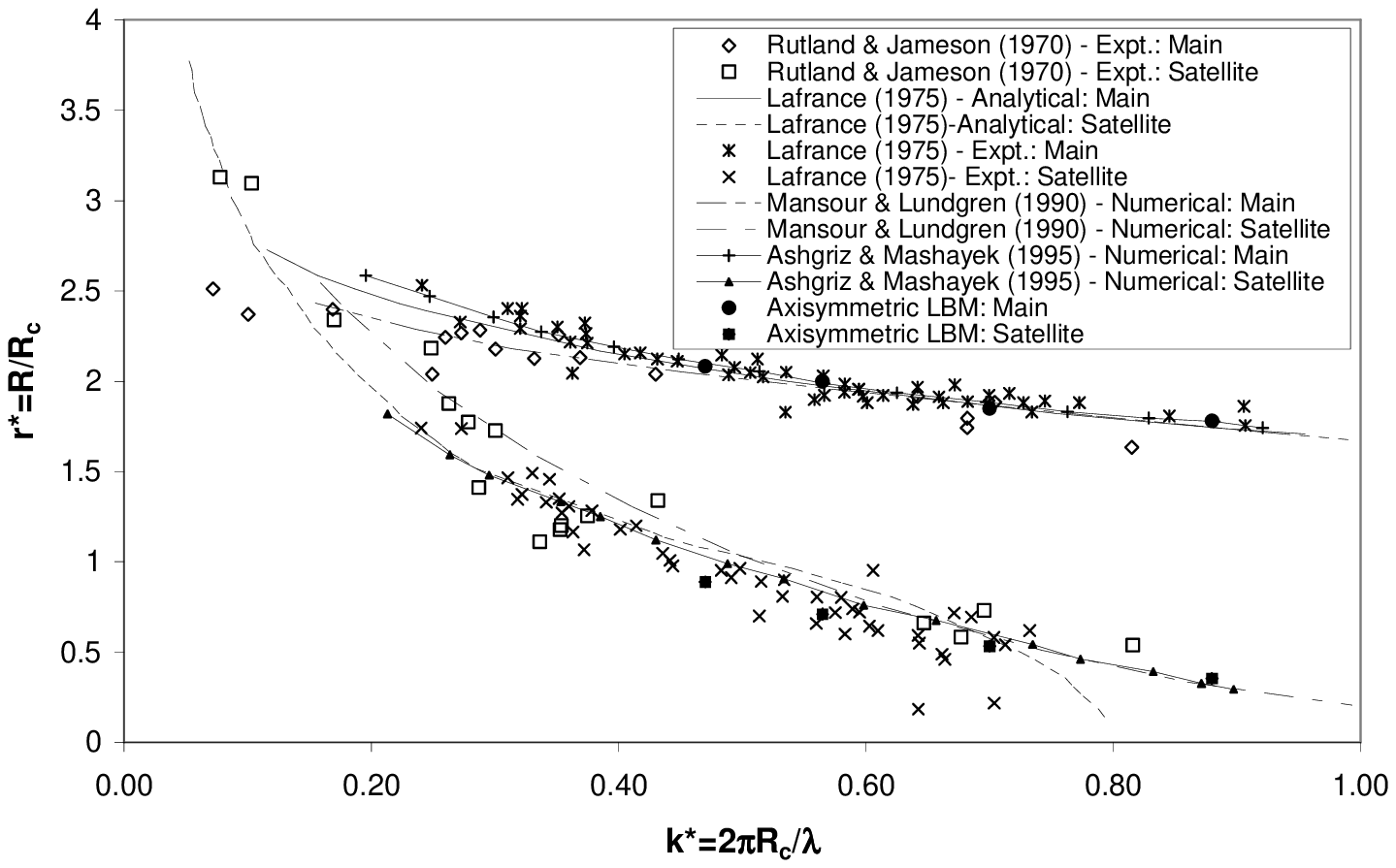}
\caption{\label{fig:rayleigh3}Drop sizes resulting from Rayleigh breakup of liquid cylindrical column as a 
	 function of wave number $k^{*}$. Quantities are dimensionless.}
\end{center}
\end{figure}

Let us now increase the wavelength of the disturbance to $\lambda_d=600$,  keeping the physical parameters the same as before.
We consider a domain represented by $601\times 151$ lattice sites. Since,
$R_c=45$ as before, the wavenumber is $0.47$. Figure \ref{fig:rayleigh2} shows the temporal evolution of the configurations of
the liquid column at this reduced wavenumber. The axisymmetric disturbance grows with time.
Since the wavelength is longer, it can be noticed that the ligament that is formed during the Rayleigh instability
is also longer. As a result, after the column breaks up, a larger satellite droplet is formed.
To express the drop size distribution with wave numbers more quantitatively, we plot the non-dimensional
size of the main and satellite drops, $r^{*}=R/R_c$, as a function of wave number, $k^{*}$ in Fig. \ref{fig:rayleigh3}. 
It may be noted that Rayleigh's
original analysis predicts only the onset of breakup and not the formation of satellite droplets. To predict analytically
satellite droplet formation, it has been shown that at least a third-order
perturbation analysis of the Navier-Stokes equations (NSE) is needed~\cite{lafrance75}. Computations based on
direct solutions of the NSE also predict the formation of the satellite droplets.

To evaluate the
drop size distribution computed using the axisymmetric LBE model, we consider the experimental data of
Rutland and Jameson (1971),~\nocite{rutland71} the experimental data and analytical solution based on a
third-order perturbation analysis of the NSE by Lafrance (1975),~\nocite{lafrance75}
a boundary integral solution of the NSE by Mansour and Lundgren (1990)~\nocite{mansour90} and a finite element solution of
the NSE by Ashgriz and Mashayek (1995).~\nocite{ashgriz95} It can be seen in the figure that as long as the wavenumber
is less than one, as expected there will be a satellite droplet formation. As the wavenumber is reduced, the sizes of
both the main drop and satellite droplet increases. The rate of increase of the size of the satellite droplet is greater
than that of the main drop. Notice that there is considerable scatter in the available data in the figure. The computed
results from the axisymmetric LBE model are presented for wavenumbers greater than or equal to $0.47$. Ignoring the
two experimental data points of Lafrance (1975) for the satellite drop sizes that deviate considerably from the others,
we find that the axisymmetric model is able to reproduce the drop size distribution quantitatively within $12\%$.

\begin{figure}
\begin{center}
\includegraphics{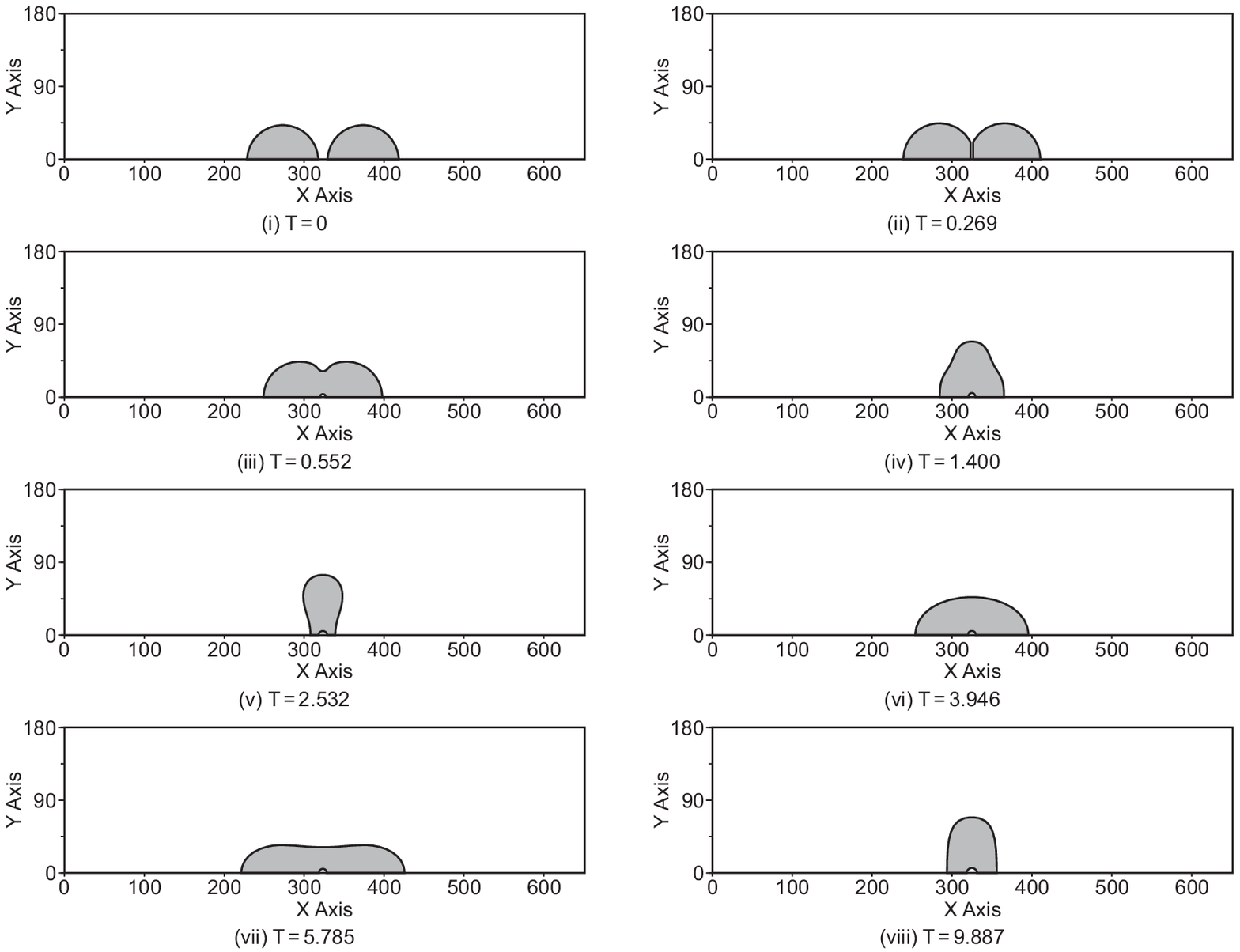}
\caption{\label{fig:collWe20} Colliding drops at different times $T$; $We=20$, $Oh=0.589$, $r=4$, $\lambda=1$. 
	 Time is normalized by the relative velocity between the drops and their diameter. Axes are in lattice units.}
\end{center}
\end{figure}

\begin{figure}
\begin{center}
\includegraphics{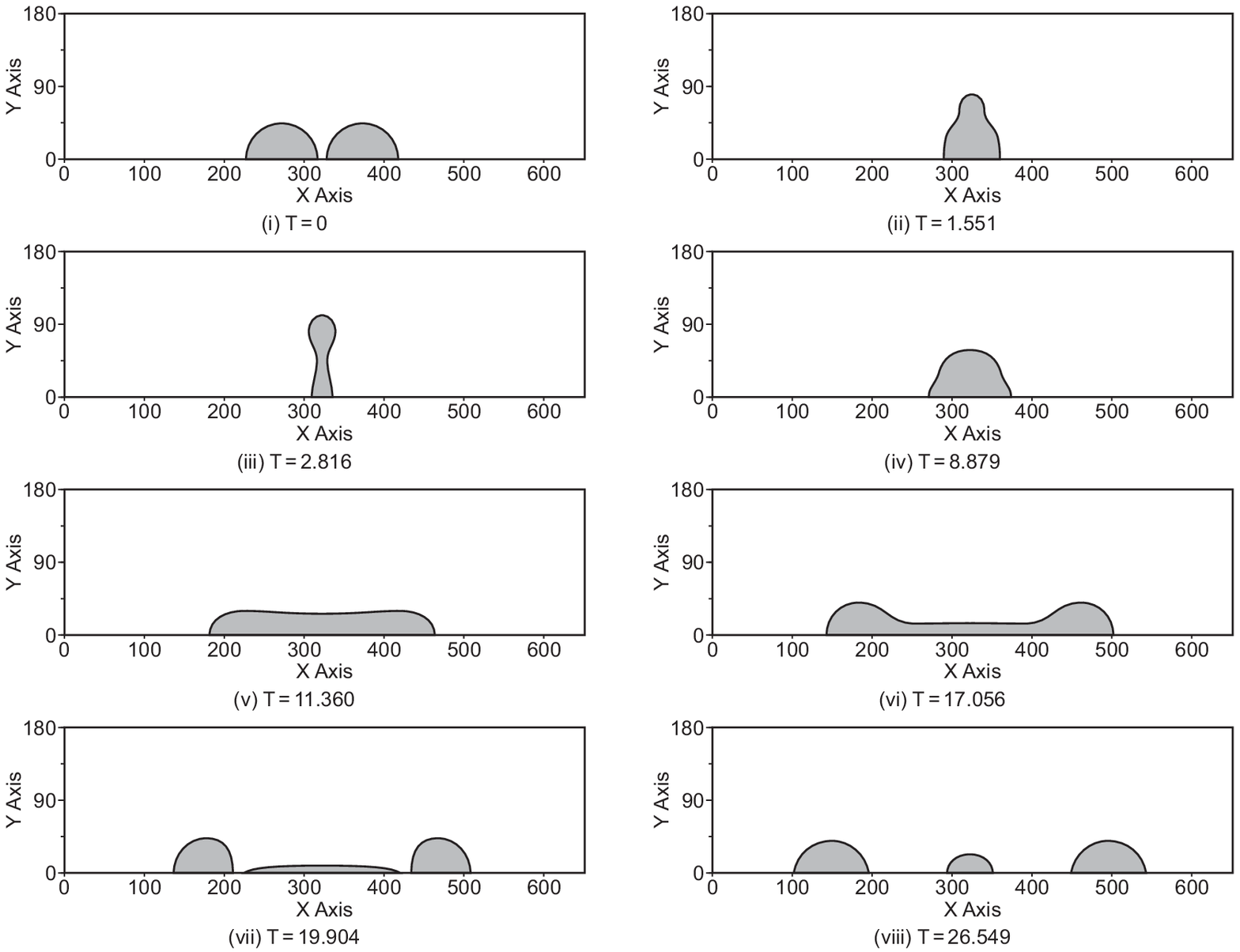}
\caption{\label{fig:collWe100} Colliding drops at different times $T$; $We=100$, $Oh=0.589$, $r=4$, $\lambda=1$. 
	 Time is normalized by the relative velocity between the drops and their diameter. Axes are in lattice units.}
\end{center}
\end{figure}

The axisymmetric model has been employed to study head-on collisions of drops of radii $R_1$ and $R_2$ approaching each other
with a relative velocity $U$. The dynamics and outcome of colliding drops is characterized mainly by the Weber number,
$We$ defined by $We=\rho_{l} (R_{1}+R_{2}) U^{2} / \sigma$~\cite{qian97}. Additional parameters that may have an influence
are the Ohnesorge number, $Oh$, defined by $Oh=16\mu_{l}/\sqrt{\rho_{l}R_{1}\sigma}$ and ratios of liquid and gas
densities($r$) and dynamic viscosities ($\lambda$). According to experiments~\cite{qian97}, it is expected that lower $We$ 
collisions
lead to coalescence while higher $We$ to separation by reflexive action. Figures \ref{fig:collWe20} and 
\ref{fig:collWe100} present drop configurations
at $We=20$ and $We=100$ respectively. Notice that at $We=20$, the drops coalesce, while at $We=100$, they eventually separate
with the formation of a satellite droplet, which are consistent with experimental observations. 
Also notice that for the latter
case, the temporarily coalesced drop undergoes various stages of deformation which are consistent with a recent theoretical
analysis~\cite{roisman04}. Additional details of these and other studies of drop collisions are given in Ref.~\cite{premnath04a}.

\section{\label{sec:summary}Summary}
In this paper, a LB model for axisymmetric multiphase flows is developed.
The axisymmetric model is developed by adding source terms to the
standard Cartesian BGK LBE. The source terms, which are temporally and
spatially dependent, represent the axisymmetric contributions of the order parameter,
which distinguish the different phases, as well as
inertial, viscous and surface tension forces. Consistency of the model in achieving the
desired axisymmetric flow multiphase behavior is established through the Chapman-Enskog
multiscale analysis.  The analysis shows that the axisymmetric macroscopic conservation
equations are recovered in the continuum limit.
An axisymmetric model with reduced compressibility effects is then
developed to improve its computational stability. In this version,
a transformation is introduced to the distribution function in the LBE such that it
reduces the compressibility effects.
Comparisons of computed axisymmetric equilibrium drop formation
and oscillations, Rayleigh capillary instability, breakup and formation of satellite
drops liquid cylindrical liquid columns and the outcomes of head-on drop collisions with 
available data show satisfactory agreement.
The maximum error for the frequency of drop oscillations is less than $5.1\%$ and that for drop sizes as a
result of Rayleigh breakup is $12\%$.

\begin{acknowledgments}
The authors thank Dr.\ X. He for helpful discussions and the Purdue University Computing Center (PUCC)
and National Center for Supercomputing Applications (NCSA) for providing access to computing resources.
\end{acknowledgments}

\bibliography{paper}

\end{document}